\documentstyle
{article}

\author{Ramil' F.Bikbaev, Vadim R.Kudashev\\
Institute of Mathematics, Chernyshevskii 112, Ufa, 450000, Russia\\
E-mail: vadkud@nkc.bashkiria.su\\
February 17, 1994}
\title{Whitham deformations partially saturating the modulational instability
in the nonlinear Schrodinger equation
}
\begin{document}

\maketitle
\begin{abstract}
In the framework of Gure\-vich and Pitaevs\-kii approach [1] we construct
modulated by Whitham [2] solution of nonlinear Shrodinger (NS) equation
partially saturating the modulational instability. This solution describes
new scenario of monochromatic wave evolution in NS equation which leads to
generation of new phase and oscillation region.
\end{abstract}

{\bf 1. }From the point of view of Whitham deformation theory [2]
modulational instability (see e.g. [3]) in the NS equation
\begin{equation}
iu_t+u_{xx}+2\mid u\mid ^2u=0
\end{equation}
is connected with the complexity (i.e. non vanishing imaginary parts) of the
characteristic speeds of the Whitham-NS equations. This leads to the
exponential growth of the perturbations in the linearized (on the locally
constant NS equation background) NS equation. It is natural to conjecture
however that there are nontrivial (no constants) solutions of Whitham-NS
system which have vanishing imaginary parts of some (not all!)
characteristic speeds and hence partially saturating the modulational
instability.

The purpose of this paper is to demonstrate on the simplest example the
existence of the approximate solution of NS equation which partially
saturates the modulational instability. This solution of (1) describes new
(in contrast with ordinary modulational instability and some of its
modifications [3-6]), nonlinear scenario of evolution of zero-phase solution
(monochromatic wave) of NS equation

\begin{equation}
u_0(x,t)=\exp (i2t),
\end{equation}
which leads to generation and propagation of new phase and oscillations in
the (1) according to Whitham-NS equations. It is important that the
characteristic development time for the new phase is the same order as the
characteristic time of development of the ordinary modulational instability.

Evolution of the new phase leads to the formation of the oscillation region
in the solution of (1). Due to the vanishing of one of the modulational
instability increments in our case an analogy appears with the known [1]
nature of the generation of new phases in modulational stable situations.
Let us stress that physically principal difference of our situation from [1]
is that the our scenario of generation of new phase which we propose below
works (in contrast with [1]) in the case when asymptotics $u(x\rightarrow
\pm \infty ,t)$ of the solution $u(x,t)$ coincide.

{\bf 2. }The approximate solution of the NS equation which we propose here
consists of the external zero-phase solution (2) in the region $x\leq
x^{-}(t)$ and $x\geq x^{+}(t)$ and from the internal modulated by Whitham
one-phase solution (see (3)-(13)) in the region $x^{-}(t)\leq x\leq x^{+}(t)$%
{}.

It is well known that one-phase solutions of (1) can be written in the form

\begin{equation}
\begin{array}{cc}
u^{\pm }=\sqrt{f(\theta ^{\pm })}\cdot \exp (i\varphi ^{\pm }), & \theta
^{\pm }=x-U^{\pm }t,
\end{array}
\end{equation}
\begin{equation}
\begin{array}{cc}
f(\theta )=f_3+(f_1-f_3)dn^2\{\sqrt{f_1-f_3}\cdot \theta ;m\}, &
m=(f_1-f_2)/(f_1-f_3),
\end{array}
\end{equation}
\begin{equation}
\begin{array}{cc}
\varphi _x^{\pm }=U^{\pm }/2\mp A/f; & \varphi _t^{\pm }=-(U^{\pm
})^2/4+(\sum\limits_if_
i)\pm U^{\pm }A/f,
\end{array}
\end{equation}
where $f_1\geq f\geq f_2\geq 0\geq f_3$, $A=\sqrt{-f_1f_2f_3}\geq 0$, $dn$ -
is the Jacobi elliptic function. Elliptic spectral curve corresponding to
(3)-(5) has branching points $(\lambda _2=\lambda _1^{*},\lambda _4=\lambda
_3^{*})$ such that (c.f. [7]):

at upper sign:

\begin{equation}
\begin{array}{c}
\lambda _1^{+}\equiv \alpha ^{+}-i\gamma ^{+}=U^{+}/4-
\sqrt{-f_3}/2-i(\sqrt{f_1}+\sqrt{f_2})/2, \\ \lambda _3^{+}\equiv \beta
^{+}-i\delta ^{+}=U^{+}/4+\sqrt{-f_3}/2-i(\sqrt{f_1}-\sqrt{f_2})/2,
\end{array}
\end{equation}

at lower sign:
\begin{equation}
\begin{array}{c}
\lambda _1^{-}\equiv \beta ^{-}-i\delta ^{-}=U^{-}/4-
\sqrt{-f_3}/2-i(\sqrt{f_1}-\sqrt{f_2})/2, \\ \lambda _3^{-}\equiv \alpha
^{-}-i\gamma ^{-}=U^{-}/4+\sqrt{-f_3}/2-i(\sqrt{f_1}+\sqrt{f_2})/2,
\end{array}
\end{equation}
Whitham-NS equations for (1), (3)-(7) are (c.f. [7])
\begin{equation}
\begin{array}{cc}
d\lambda _i/dt+S_i(\lambda )d\lambda _i/dx=0, & i=1,2,3,4,
\end{array}
\end{equation}
\begin{equation}
\begin{array}{ll}
S_1=U+2\lambda _{12}/(1-\mu \lambda _{32}/\lambda _{31}), & S_3=U+2\lambda
_{34}/(1-\mu \lambda _{14}/\lambda _{13}), \\
S_2=S_1^{*},S_4=S_3^{*}, & \mu \equiv E(m)/K(m),
\end{array}
\end{equation}
where $\lambda _{ij}\equiv \lambda _i-\lambda _j$, $m=\lambda _{21}\lambda
_{43}/\lambda _{32}\lambda _{14}$, $E,K$ - are the complete elliptic
integrals of the second and first kind respectively.

In the oscillation region $0\leq x\leq x^{+}(t)$, we use special solution of
the Whitham-NS system (8)
\begin{equation}
\begin{array}{cc}
\lambda _1^{+}\equiv \alpha ^{+}-i\gamma ^{+}\equiv const.; & Im(S_3)=0, \\
\{4\beta ^{+}+2[(\gamma ^{+})^2-(\delta ^{+})^2]/(\beta ^{+}-\alpha
^{+})\}t-x=g(\beta ^{+},\delta ^{+}), &
\end{array}
\end{equation}
where $g(\beta ,\delta )$ - is an arbitrary smooth function of its
variables, which is determined from the initial conditions. Let us consider
the simplest case $g\equiv 0$ (c.f. [1,8]). From the initial condition (2)
we get $\gamma ^{+}\equiv 1$, $\alpha ^{+}\equiv 0$, and system (10) are
invariant with respect to transformations $\delta ^{+}\rightarrow -\delta
^{+}$ and $(\beta ^{+},x)\rightarrow -(\beta ^{+},x)$. Additional analysis
shows that the system (10) is compatible and has unique solution with $%
\delta ^{+}\geq 0$, $\beta ^{+}\geq 0$, in the region $0\leq x\leq x^{+}(t).$
Near the boundary $x^{+}=x^{+}(t)$ the solution of the system (10) has form
\begin{equation}
\begin{array}{ccc}
x^{+}=4\sqrt{2}t, & x=x^{+}-x^{\prime }, & 0\prec x^{\prime }\ll 1, \\
\beta ^{+}\approx 1/\sqrt{2}-7x^{\prime }/48t, & (\delta ^{+})^2\approx
x^{\prime }/2\sqrt{2}t. &
\end{array}
\end{equation}
At the boundary $x=x^{+}(t)$ the solution $u^{+}$ from (3)-(5) is
continuously glued with $u_0$ from (2). If $(x/t)\rightarrow +0$ the points $%
(\lambda _3^{+},\lambda _4^{+})$ closely come to the points $(\lambda
_1^{+},\lambda _2^{+}).$ In this limit our solution (3) degenerates into the
soliton.

In the oscillation region $x^{-}(t)\leq x\leq 0$ solution of system (8) is
defined by equations

\begin{equation}
\begin{array}{cc}
\lambda _3^{-}\equiv \alpha ^{-}-i\gamma ^{-}\equiv const.; & Im(S_1)=0, \\
\{4\beta ^{-}+2[(\gamma ^{-})^2-(\delta ^{-})^2]/(\beta ^{-}-\alpha
^{-})\}t-x=0. &
\end{array}
\end{equation}
System (12) is analogous to system (10). From the initial condition (2) we
obtain $\gamma ^{-}\equiv 1$, $\alpha ^{-}\equiv 0.$ Near the boundary $%
x^{-}=x^{-}(t)$ solution of system (12) has form
\begin{equation}
\begin{array}{ccc}
x^{-}=-4\sqrt{2}t, & x=x^{-}+x^{\prime \prime }, & 0\prec x^{\prime \prime
}\ll 1, \\
\beta ^{-}\approx -1/\sqrt{2}+7x^{\prime \prime }/48t, & (\delta
^{-})^2\approx x^{\prime \prime }/2\sqrt{2}t. &
\end{array}
\end{equation}
At the boundary $x=x^{-}(t)$ the solution $u^{-}$ from (3)-(5) is
continuously glued with $u_0$ from (2). If $(x/t)\rightarrow -0$ the points $%
(\lambda _1^{-},\lambda _2^{-})$ closely come to the points $(\lambda
_3^{-},\lambda _4^{-}).$ In this limit our solution (3) degenerates into the
soliton.

Due to Galilean invariance of (1) the above analysis may be easily extended
to the case of the zero-phase solution (monochromatic wave)

$$
u_0(x,t)=\exp [i2\alpha x+i(2-4\alpha ^2)t].
$$
Note the corresponding changes in the formulae (10)-(13): $x^{\pm
}\rightarrow 4\alpha t+x^{\pm }$, $\alpha ^{\pm }\rightarrow \alpha $, $%
\beta ^{\pm }\rightarrow \alpha +\beta ^{\pm }$, $\lambda _1^{+}\rightarrow
\alpha +\lambda _1^{+}$, $\lambda _3^{-}\rightarrow \alpha +\lambda _3^{-}.$

{\bf 3. }One - phase Whitham-NS equations (8) describe generally speaking
two pairs of perturbations (corresponding to 4 branching points of our
elliptic curve). It seems natural to ask is it correct to ''forget ''
about one pair of perturbations (note that we imposed restriction $\lambda
_1^{+}\equiv const$. or $\lambda _3^{-}\equiv const$.) and investigate the
above process in the ''pure form''? The answer is that due to (11), (13) $%
\mid dx^{\pm }(t)/dt\mid \cong 1$, hence the characteristic time of the
process considered above is of the same order as the characteristic time of
development of modulational instability of the ''forgotten ''(''thrown'')
mode $((\mid ImS_1^{+}\mid ,\mid ImS_3^{-}\mid )\cong 1$ at the initial
time. Let us note also that $(\mid ImS_1^{+}\mid ,\mid ImS_3^{-}\mid )$ have
maximal values in the external region and decrease (up to zero) by
approaching to the center of the oscillation region.

Our consideration in this work are not rigorous. They are performed on the
level of Gurevich and Pitaevskii [1] reasoning. Rigorous analysis of the
Cauchy problem for (1), (2) requires much more sophisticated technique. One
can get the impression of the corresponding difficulties on the simplest
example of modulationally stable situation by comparing [1] with the papers
[9] where justification of Gurevich and Pitaevskii conjecture (in the case
of step-like initial data) was done.

{\bf Acknowlegment}

The work of V.R.K. was supported, in part, by ''BUT-RUTEX '', and by RFFI
grant 93-011-16088.

{\bf References}

[1] A.V.Gurevich and L.P.Pitaevskii, JETP 38 (1974) 291.

[2] G.B.Whitham, Proc. Roy. Soc. A283 (1965) 238; Linear and Nonlinear
waves, John Willey and sons, 1974, N.Y.

[3] B.B.Kadomtsev, Collective phenomena in plasmas, Moscow, Nauka, 1976 [in
Russian].

[4] E.A.Kuznetsov, Dokl. Akad. Nauk SSSR, 236 (1977) 575; N.N.Ahmediev,
V.M.Eleonskii and N.E.Kulagin, Zh. Eksp. Teor. Fiz. 89 (1985) 1542.

[5] S.K.Zhda\-nov and B.A.Trub\-ni\-kov, Quasi-ga\-seous un\-stable media,
Mos\-cow,
Nau\-ka, 1991 [in Russian].

[6] G.Rowlands, J. Phys.A: Math. Gen. 13 (1980) 2395; P.A.E. Janssen, Phys.
Fluids 24 (1981) 23; E. Infeld, Phys. Rev. Lett. 47 (1981) 717.

[7] M.V.Pavlov, Teor. Mat. Fiz., 71 (1987) 351.

[8] A.M.Kamchatnov, Phys. Lett., 162A (1992) 389.

[9] R.F.Bikbaev, Teor. Mat. Fiz., 81 (1989) 3; P.D.Lax, C.D.Levermore and
S.Venakides, Generation and propagation of oscillations in dispersive IVP's
and their limiting behavior, Important developments in Soliton Theory,
1980-1990, Springer-Verlag, Berlin, 1993 (T.Fokas, V.E.Zakharov eds.).

\end{document}